\documentclass[
reprint,
longbibliography,
amsmath,amssymb,
aps,
prl,
floatfix,
]{revtex4-2}

\usepackage{graphicx}%
\usepackage{xcolor}
\usepackage{cancel}
\usepackage{dcolumn}%
\usepackage{bm}%
\usepackage[colorlinks, linkcolor=blue, citecolor=blue, urlcolor=blue]{hyperref}%
\usepackage{booktabs}

\begin{document}

\title{Different fractional charges from auto- and cross-correlation noise in quantum Hall states without upstream modes}
\author{Navketan Batra}
\author{D. E. Feldman}
\affiliation{Department of Physics, Brown University, Providence, Rhode Island 02912, USA}
\affiliation{Brown Theoretical Physics Center, Brown University, Providence, Rhode Island 02912, USA}

\date{\today}

\begin{abstract}
Fractional charges of anyons can be extracted from shot noise in two ways. One can use either the auto-correlation noise of the current in one drain or the cross-correlation noise between two drains on the two sides of the device. The former approach typically overestimates the charge.
This may happen due to upstream edge modes. We propose a mechanism for the excess auto-correlation noise
without upstream modes. It applies to systems with multiple co-propagating edge modes and assumes that the noise is measured at a low but non-zero frequency.

\end{abstract}

\maketitle

\begin{figure}
    \centering
    \includegraphics[scale=1.1]{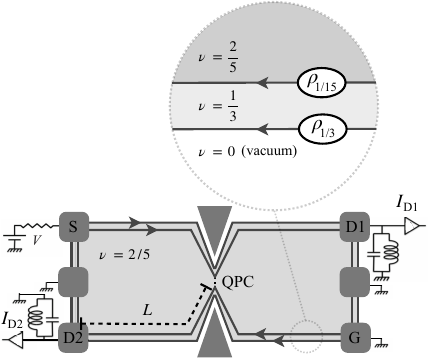}
    \caption{Current-noise measurement setup for quasiparticle tunneling across a quantum point contact (QPC). The edge structure of a $\nu=2/5$ FQH state comprises of two downstream modes separated by a $\nu=1/3$ incompressible state. Current flows into the sample through the Ohmic contact S to impinge on the QPC. Due to charge partitioning, shot noise is generated at the QPC. The noise is evaluated by measuring the current at drains D1 and D2.}
    \label{fig:1}
\end{figure}

Fractional charge has been recognized as a key feature of the fractional quantum Hall effect (FQHE) since the early days of the field \cite{laughlin:1983,feldman:2021}.  The most successful experimental technique to probe fractional charges uses shot noise \cite{feldman:2021,dePicciotto:1997,saminadayar:1997,reznikov:1999,dolev:2008,jos,heiblum:2020}.
It helped observe anyons of charges $e/3$, $e/4$, $e/5$, and $e/7$. In addition, noise is a sensitive measure of anyon statistics \cite{safi:2001,kim:2006,rosenow:2016,bartolomei:2020,glidic:2023}
and can be used to distinguish between different $\nu=5/2$ candidate states \cite{manna:2023}. The technique is implemented by creating a point contact between two FQHE edges (Fig. \ref{fig:1}) and measuring the low-frequency current noise at an angular frequency $\omega\sim 10^7$ Hz. At a low current, the noise is expected to be proportional to the charge of the tunneling quasiparticle.

This agrees with experiments at higher temperatures $T\sim 100$ mK. At low temperatures, $T\sim 10$ mK, the observed charge is often higher than the anyon charge predicted by theory \cite{chung:2003,private,feldman:2021,glidic:2023}. This typically happens in states with topologically protected upstream edge modes, such as the $\nu=2/3$ \cite{biswas:2022,private} and $\nu=3/5$ 
states \cite{feldman:2017}. The discrepancy between theory and experiment may be explained by the effect of upstream modes excited at the tunneling contact in tunneling events \cite{biswas:2022,private}. 
They carry energy back to the current source and generate additional noise by heating the source. This explanation was tested by comparing the auto-correlation noise in one drain (Fig. \ref{fig:1}) with the cross-correlation noise between two drains \cite{private} at $\nu=2/3$. In agreement with the above physical picture, excess noise was only seen in auto-correlations of the current, and a correct quasiparticle charge was extracted from the cross-correlation noise.

Explaining the observed excess shot noise \cite{chung:2003} at the filling factor $2/5$ is harder since no topologically protected upstream modes exist at that filling factor. Nevertheless, the same phenomenology \cite{private} was observed at that filling factor as at $\nu=2/3$. Non-topological modes may emerge from edge reconstruction at any filling factor \cite{reconstruction}. However, they decay rapidly with the distance and were observed on shorter scales \cite{rec1,rec2} than the propagation distance $\sim 50$ \textmu m on the $2/5$ edge \cite{private}. Besides, similar edge-reconstruction physics is expected at the nearby filling factor $1/3$, where excess noise was not observed in either cross- or auto-correlation experiment \cite{private,glidic:2023}. It is thus of interest to find a mechanism of excess-noise generation in the absence of upstream modes. We propose such a mechanism in this paper.

Three ingredients are essential for the mechanism. First, we need two or more downstream modes of different velocities. Second, the mechanism requires tunneling between downstream channels but assumes that, with a possible exception of the region near the tunneling contact, the equilibration length between the channels is longer than the thermal length. Finally, it is crucial that the experiment is performed at a low but non-zero frequency $\omega$. These conditions are consistent with the filling factor $\nu=2/5$ but do not apply at $\nu=1/3$, where only one downstream mode exists. They also do not apply at $\nu=2$, where spin conservation suppresses tunneling between the two edge channels with the up and down spin orientations, and where no experimental evidence of excess noise has been reported.

With the above ingredients  present, the tunneling current through the point contact generates Joule heat. This results in excess thermal noise in the downstream channels. The effect is only present with more than one channel since otherwise, charge conservation blocks noise generation. With more than one channel, charge conservation ensures that the sum of the heat-induced fluctuating charge currents in all channels remains zero. 
At zero frequency, this would be the end of the story. At a low but non-zero frequency, we still expect that the heat-induced charge current densities sum to zero over all channels near the point contact at any moment in time. Yet, the sum is non-zero near the drain due to different propagation times along the channels. This explains excess auto-correlation noise. Since heat-induced currents are uncorrelated on the two sides of the point contact, no excess cross-correlation noise is generated. 

Below, we start with a discussion of the normal edge modes at $\nu=2/5$. We then estimate the magnitude of the heating effect and introduce a model for the calculation of noise. We finish with numerical estimates and a discussion of possible implications and experimental tests of the mechanism.

\begin{figure}
    \centering
    \includegraphics[scale=1.1]{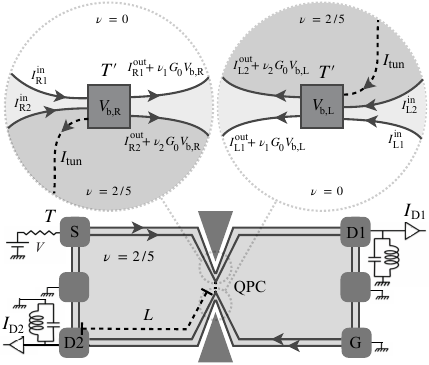}
    \caption{Schematics of the black box. We define $G_0=e^2/h$. 
    The Joule heat, generated due to quasiparticle tunneling through the QPC, dissipates in the black box and raises its temperature to $T'$. In order to maintain charge neutrality across the region, the voltage $V_{\text{b}}$ of the black box fluctuates, giving rise to fluctuating normal-mode charge currents. 
    }
    \label{fig:2}
\end{figure}

The $2/5$ state \cite{WenBook} is a daughter state of the Laughlin state at $\nu=1/3$.
It has two interacting co-propagating edge modes. The outer edge channel separates filling factors $0$ and $1/3$ and carries the linear charge density 
$\rho_{1/3}=e\partial_x\phi_{1/3}/2\pi$, where $\phi_{1/3}$ is a Bose field.
The inner channel separates $\nu=1/3$ from $\nu=2/5$ and carries the linear charge density $\rho_{1/15}=e\partial_x\phi_{1/15}/2\pi$. Quasiparticles of charge
$e/5$ tunnel between the inner channels of the upper and lower edge at the point contact, Fig. \ref{fig:1}. There are two equally relevant quasiparticle operators:
${\Psi_{1/5}=}\exp(3i\phi_{1/15})$ and ${\Psi=\Psi_{1/5}\Psi_{1/3}=}\exp(i\phi_{1/3}-2i\phi_{1/15})$. The former operator creates a quasiparticle on the inner edge, while the latter creates a quasiparticle of charge $e/5$ on the inner edge {via the action of $\Psi_{1/5}$} and moves charge $e/3$ between the outer and inner edge channels { via the action of
$\Psi_{1/3}=\exp(i\phi_{1/3}-5i\phi_{1/15})$
}. 

It is convenient to switch to the charge mode $\phi_c=\phi_{1/3}+\phi_{1/15}$ and the neutral mode $\phi_n=\phi_{1/3}-5\phi_{1/15}$. The charge current 
$j(x,t)=e\partial_t\phi_c/2\pi$. The Lagrangian density
on the left-moving lower edge is
  \begin{align}
  \label{1}
 \mathcal{L} &= \frac{1}{4\pi}\bigg[  \frac{5}{2}\partial_t\phi_{c}\partial_x\phi_{c} + \frac{1}{2} \partial_t\phi_{n}\partial_x\phi_{n} - \frac{5}{2}v_{c}(\partial_x\phi_{c})^2 \nonumber \\
 &~~~~~~~~~~~~~ - \frac{1}{2} v_{n} (\partial_x\phi_{n})^2 
 - 2\Delta_{}\partial_x\phi_{c}\partial_x\phi_{n}   \bigg],
\end{align}
where $v_c$ and $v_n$ are the mode velocities and $\Delta$ is the inter-mode interaction. We expect that \cite{velocities} $v_c>v_n\sim\Delta$.

Next, we introduce the normal modes $\phi_{1,2}$ defined as
\begin{equation}
\label{2}
\phi_i=\left(\sqrt{5}\xi^1_i(\gamma) \phi_c + \xi^2_i(\gamma) \phi_n\right)/\sqrt{2};~~~~ (i=1,2),
\end{equation}
where $\xi^{1}_1(\gamma)= -\xi_2^2(\gamma)\equiv\cos(\gamma)$, $\xi^{1}_2(\gamma)= \xi^2_1(\gamma)\equiv\sin(\gamma)$, and
\begin{equation}
\label{4}
\tan 2\gamma=\frac{4\Delta}{\sqrt{5}(v_c-v_n)}.    
\end{equation} 
The normal modes diagonalize the Hamiltonian and commute with each other so that the Lagrangian density becomes
\begin{equation}
\label{5}
\mathcal{L}=\frac{1}{4\pi}[\partial_t\phi_1\partial_x\phi_1+\partial_t\phi_2\partial_x\phi_2-
v_1(\partial_x\phi_1)^2-v_2(\partial_x\phi_2)^2],
\end{equation}
where $v_1 = (v_c\cos^2\gamma - v_n\sin^2\gamma)/\cos(2\gamma)$ and $v_2 = (v_n\cos^2\gamma - v_c\sin^2\gamma)/\cos(2\gamma)$. Both normal modes carry charge since $\phi_c=\sqrt{\nu_1}\phi_1+\sqrt{\nu_2}\phi_2$, where 
$\nu_1=(2\cos^2\gamma)/5$, $\nu_2=(2\sin^2\gamma)/5$. 
Hence, the total charge current $I(x,t)$ reduces to the sum of the currents in the normal modes,
\begin{equation}
\label{6}
I=I_1+I_2=\frac{\sqrt{\nu_1}e\partial_t\phi_1}{2\pi}
+\frac{\sqrt{\nu_2}e\partial_t\phi_2}{2\pi}.
\end{equation}
The response of the normal modes to a voltage bias $V$
can be computed by observing that in thermal equilibrium, 
$V$ is the chemical potential conjugated to the electric charge. Hence, the chemical potential, conjugated to a normal mode $\phi_i$, is $\sqrt{\nu_i}eV/2\pi$. We then find
$I_i=\nu_i e^2 V/h$. We see that $\{\nu_i\}$ play the role of the filling factors associated with the normal modes. In particular, $\nu_1+\nu_2=2/5$.

Note a significant difference from the $\nu=2/3$ state and other hole-like quantum Hall states, where the normal modes are just the charge and neutral modes \cite{WenBook}. Such states are daughter states of a liquid with a higher filling factor, and electrostatics force contra-propagating modes to run close to each other \cite{CSG}. Hence, inter-mode tunneling dominates short-range physics. As a result, the charge equilibration length \cite{eqv} is very short $\lesssim 1$ \textmu m.
At $\nu=2/5$, the distance between the inner and outer modes is greater than at $\nu=2/3$, and inter-mode tunneling has little effect on the normal modes.

Since the region near the point contact differs from the rest of the edge, one can imagine two scenarios for charge tunneling between the
inner and outer modes. Tunneling strength may be moderate and comparable at all points along the edge. Alternatively, inter-mode transport may be dominated by strong tunneling in the region near the point contact. We expect similar physics in both cases and will focus on the latter scenario, which allows simpler 
calculations. In this scenario, the above results for normal modes do not apply to a region near the contact. We will call it the black box  (Fig. \ref{fig:2}). The black box is the place where Joule heat is dissipated. 

While we focus on the black box model due to its simplicity, the  experimental results 
\cite{glidic:2023} at $\nu=2/5$ point at the scenario with moderate tunneling along the edge. Indeed,
Fig. 11 of Ref. \onlinecite{glidic:2023} shows that about 5\% of the current is exchanged between edge channels on the scale of 1 \textmu m. This suggests significant charge and energy exchange between the two modes on the scale of the total edge length. Note that thermalization in each of the two channels may happen on a shorter scale due to disorder-induced tunneling between different segments of the edge, compressible puddles, and channels present on short scales due to edge reconstruction \cite{FH2022}. 
One might also need to consider electron-phonon interaction \cite{ep}. Excess noise of the same order of magnitude is expected in this picture and in a simpler black-box model.

Quantitative analysis of non-equilibrium physics in the black box is challenging. Thus, we will only roughly estimate the excess noise. We will do so by assuming that the black box is at equilibrium at the temperature determined by Joule heating. We estimate the dissipated heat on each side of the point contact as $I_{\text{{tun}}}V/2$, where $V$ is the voltage bias, $I_{\text{{tun}}}=rGV$ is the tunneling current, the Hall conductance $G=\nu e^2/h$, $\nu=2/5$, and $r\ll 1$ is the transmission of the contact.
Next, we make use of the quantized heat conductance \cite{pendry:1983} of a two-channel edge $\kappa(T)=2\kappa_0(T)=2\pi^2T/3h$, and write the heat balance equation
\begin{equation}
\label{7}
\frac{\kappa(T')T'}{2}=\frac{I_{\text{{tun}}}V}{2}+\frac{\kappa(T)T}{2},
\end{equation}
where $T'$ is the temperature of the black box, and $T$ is the temperature of the sources.
Hence,
\begin{equation}
\label{8}
T'=\sqrt{r\frac{3e^2V^2}{5\pi^2}+T^2}.
\end{equation}

The heating effect generates excess current noise. We focus on the lower edge.
We will first ignore charge conservation in addressing the noise. Equation (\ref{5}) has the same structure as the Lagrangian density on the edge of the $\nu=2$ state. 
At a low frequency, the equilibrium fluctuations of the currents $\tilde I^{\text{out}}_i=\partial_t\phi_i/2\pi$ of the fields $\partial_x\phi_i/2\pi$ on the left of the black box are given by the Nyquist formula
$\langle \{\tilde I^{\text{out}}_i(\omega),\tilde I^{\text{out}}_j(-\omega)\}\rangle=\delta_{ij}\frac{2T'}{h},$
where the curly brackets denote an anticommutator. The associated fluctuating charge currents $I^{\text{out}}_i=e\sqrt{\nu_i}\tilde I^{\text{out}}_i$ obey the relation
\begin{equation}
\label{9}
\langle \{I^{\text{out}}_i(\omega),I^{\text{out}}_j(-\omega)\}\rangle=\delta_{ij}\frac{2\nu_i e^2T'}{h}.
\end{equation}

The total charge current, resulting from such Nyquist fluctuations, differs from the sum of the tunneling current $I_{\text{{tun}}}$ and the current $I^{\text{in}}$ arriving into the black box along the edge. Neglecting the capacitance of the small black box, we arrive at a contradiction with charge conservation. To resolve it, we have to remember that the voltage $V_{\text{b}}$ of the box fluctuates to ensure its charge neutrality:
\begin{equation}
\label{10}
I_1^{\text{out}}+I_2^{\text{out}}+GV_{\text{b}}(t)=I^{\text{in}}+I_\text{{tun}},
\end{equation}
where we use instantaneous values of the charge currents near the location of the box and
$I^{\text{in}}$ satisfies the Nyquist relation 
$\langle \{ I^{\text{in}}(\omega),I^{\text{in}}(-\omega)\}\rangle=2GT$. From Eq. (\ref{10}) we find $V_{\text{b}}(t)$ and then the fluctuating charge currents in the normal modes $I_i(x=0)=I_i^{\text{out}}+\frac{\nu_ie^2}{h}V_{\text{b}}$ near the location of the black box,
\begin{align}
\label{11}
I_{1} &= \frac{\nu_1}{\nu} I^{\text{in}} + \frac{\nu_2}{\nu} I^{\text{out}}_{1} - \frac{\nu_1}{\nu} I^{\text{out}}_{2} + \frac{\nu_1}{\nu} I_{\text{{tun}}}, \\
\label{12}
I_{2} &= \frac{\nu_2}{\nu} I^{\text{in}} + \frac{\nu_1}{\nu} I^{\text{out}}_{2} - \frac{\nu_2}{\nu} I^{\text{out}}_{1} + \frac{\nu_2}{{\nu}} I_{\text{{tun}}}.
\end{align}
The total charge current immediately to the left of the box is $I(0,t)=I_1(0,t)+I_2(0,t)$.
One can see that $I^{\text{out}}_i$ drops out from the above sum, which is thus insensitive to $T'$ and Joule heating.

Consider now the charge current near the drain at the distance $L$ from the point contact, where $I(-L,t)=I_1(0,t-L/v_1)+I_2(0,t-L/v_2)$. We introduce the lag time $\tau=L/v_2-L/v_1$, and define the auto-correlation noise as $S_{\text{auto}}=\langle \{I(-L,\omega),I(-L,-\omega)\}\rangle $. We observe, 
\begin{align}
\label{13}
S_{\text{auto}} =&  \langle
\{I_1(0,\omega),I_1(0,-\omega)\}\rangle + \langle \{I_2(0,\omega),I_2(0,-\omega)\}\rangle \nonumber \\
&+\exp(-i\omega\tau)\langle\{I_1(0,\omega), I_2(0,-\omega)\}\rangle
\nonumber\\
&+\exp(i\omega\tau)\langle \{I_2(0,\omega),I_1(0,-\omega)\}\rangle.
\end{align}
The experimentally observed noise also contains the contribution
$2GT$ due to the equilibrium Nyquist noise on the left vertical edge in Fig. \ref{fig:1}. We will ignore it since we focus on excess non-equilibrium noise.
We substitute Eqs. (\ref{11},\ref{12}) into the expression above for the auto-correlation noise.

Due to chiral transport, the heating process occurs downstream, that is, to the left of the tunneling contact. Hence the auto-correlation function of $I_{\text{{tun}}}$ and its cross-correlation function with $I^{\text{in}}$ do not depend on $T'$. They can be found from detailed balance \cite{levitov:2004,feldman:2017} and a fluctuation-dissipation theorem \cite{fdt1,fdt2}: 
\begin{eqnarray}
\label{14}
\langle\{I_{\text{{tun}}}(\omega),I_{\text{{tun}}}(-\omega)\}\rangle=
2e^*\langle I_{\text{{tun}}}\rangle\coth\left(\frac{e^*V}{2T}\right); \\
\label{15}
\langle\{I_{\text{{tun}}}(\pm \omega),I^{\text{in}}(\mp \omega)\}\rangle=
-2T\frac{\partial\langle I_{\text{{tun}}}\rangle}{\partial V},
\end{eqnarray}
where $e^*=e/5$ is the charge of the tunneling quasiparticle.
We combine the above expressions into
\begin{equation}
\label{16}
S_{\text{shot}}=2e^*\langle I_{\text{{tun}}}\rangle\coth\left(\frac{e^*V}{2T}\right)
-4T\frac{\partial\langle I_{\text{{tun}}}\rangle}{\partial V}.
\end{equation}
We expect no correlation between $I^{\text{out}}_i$ and the currents
$I_{\text{{tun}}}$ and $I^{\text{in}}$, and find the auto-correlation noise
\begin{equation}
\label{17}
S_{\text{auto}}=S_{\text{N}}+S_{\text{shot}}+S_{\text{ex}},
\end{equation}
where $S_{\text{N}}=2GT$ is the Nyquist contribution, and the excess noise
\begin{equation}
\label{18}
S_{\text{ex}}=\frac{4\nu_1\nu_2}{\nu^2}\sin^2(\omega\tau/2)[2G(T'-T)
-S_{\text{shot}}].
\end{equation}
The above equation explains why the excess noise is stronger at lower temperatures. Indeed, at a high temperature, $T'-T\sim reV^2/T\ll\sqrt{r}eV$, while at $T\rightarrow 0$, $T'\sim \sqrt{r}eV$. 

What about the cross-correlation noise? To compute it, one needs to derive analogs of Eqs. (\ref{11},\ref{12}) on the upper edge.
No correlation exists between the fields $I^{\text{in}}$ and $I^{\text{out}}$ on the upper and lower edges. Nevertheless, the cross-correlation noise has a correction due to the different velocities of the normal modes. The reason consists in the splitting of the tunneling current into the charge currents in the normal modes. Assuming the same length $L$ between the point contact and the drains on the upper and lower edges, there is an identical time lag $\tau$ on both edges between the tunneling charges arriving at the drains in the two modes. The cross-correlation  noise is defined as
\begin{align}
\label{19}
S_{\text{cross}}=\frac{1}{2}\Big[&\langle\{I_{\text{L}}(-L,\omega),I_{\text{R}}(L,-\omega)\}\rangle \nonumber \\
&+\langle\{I_{\text{L}}(-L,-\omega),I_{\text{R}}(L,\omega)\}\rangle\Big],
\end{align}
where $I_{\text{L,R}}$ are the drain currents on the left-moving and right-moving edges (see Fig. \ref{fig:2}). We find
\begin{equation}
\label{20a}
S_{\text{cross}}=-S_{\text{shot}}\left[1-\frac{4\nu_1\nu_2}{\nu^2}\sin^2(\omega\tau/2)\right].
\end{equation}
We will see that the frequency-dependent correction to the noise (\ref{20a}) is an order of magnitude lower than in the auto-correlation noise. This explains why the cross-correlation noise gives a more accurate value of the fractional charge.

We will now focus on the zero-temperature limit, where the excess noise is most important. Equation (\ref{17}) simplifies to
\begin{equation}
\label{21}
S_{\text{auto}}=S_{\text{shot}}+\frac{4\nu_1\nu_2}{\nu^2}\sin^2(\omega\tau/2)\left(2GT'
-S_{\text{shot}}\right).
\end{equation}
Shot noise experiments are performed at low transmission $r$. Setting $r=0.01$, which corresponds to the 6\% transmission of the inner channel, one finds
\begin{equation}
\label{21a}
S_{\text{auto}}=S_{\text{shot}}+11.328\times\frac{4\nu_1\nu_2}{\nu^2}\sin^2(\omega\tau/2)S_{\text{shot}},
\end{equation}
where we used Eq. (\ref{8}) to find 
$T'={\rm const}\times  V\sqrt{r}$.

At zero temperature, the Fano factor is defined as $F=S_{\text{auto}}/2eI_{\text{tun}}$ and reduces to $e^*/e$ without excess noise. It is expected to be opposite to the combination $S_{\text{cross}}/2eI_{\text{tun}}$ at zero temperature. Experimentally, $S_{\text{cross}}+S_{\text{auto}}$ differs significantly from 0.
The correction to the Fano factor
\begin{equation}
\label{22}
\Delta F=\frac{S_{\text{auto}}+S_{\text{cross}}}{2eI_{\text{tun}}}=\frac{e^*}{e}\times 12.328\times\frac{4\nu_1\nu_2}{\nu^2}\sin^2(\omega\tau/2).
\end{equation}

We now proceed to numerical estimates. We chose $v_c\sim \nu e^2/(\epsilon h)\sim 5\times 10^6$ cm$/$s, where $\epsilon$ is the dielectric constant. Neutral mode velocity $v_n$ is expected to be lower than the charge mode velocity $v_c$,
and we set $v_n=10^6$ cm$/$s. We assume $L=100$ \textmu m and 
$\omega=10^7$ Hz. The intermode interaction $\Delta$ is expected to be comparable to $v_n$. Stability of the theory (\ref{1}) requires 
$|\Delta|<\sqrt{5v_cv_n}/2=2.5\times 10^6$ cm$/$s.

The effect is weak when $|\Delta|<v_n$. At $|\Delta|=2\times 10^6$ cm$/$s, one finds $\Delta F/F\approx 0.12$. At $|\Delta|=2.25\times 10^6$ cm$/$s, one finds $\Delta F/F\approx 0.54$. Excess noise becomes very large if $\Delta$
approach the limit of stability, but there is no obvious reason for it to do so, and the observed $\Delta F<F$. It should be remembered that our model allows only a crude estimate of the excess noise. It is likely that the magnitude of the effect is sample sensitive.

In our model, we assume position-independent velocities along the edge. The choice of $\Delta_{}>2v_n$ corresponds to a relatively weak interaction between the inner and outer modes $\phi_{1/15}$ and $\phi_{1/3}$. Such interaction is indeed screened on a gate-defined edge. Actual samples combine gate-defined and etched sections. Hence, the edge velocities are position-dependent. The lag time is dominated by the sections with the lowest velocity. Interaction between contra-propagating edge segments is possible across narrow gates and decreases edge velocities \cite{v1,v2}.

Our mechanism involves charge transport along both inner and outer channels. Hence, the mechanism can be tested by probing currents in both channels. In the scenario with strong inter-channel coupling near the point contact, the effect becomes stronger with the increasing edge length and exhibits a characteristic $\sin^2(\omega\tau/2)$ frequency dependence.
For short edges, the dominant source of excess noise is likely edge reconstruction. For intermediate lengths, edge reconstruction may coexist with our mechanism, which becomes dominant for long edges. In a version of our mechanism, the time lag $\tau$ comes from a time difference for the absorption of the inner and outer modes by drains.

We have only considered charge currents above, but the total current includes the displacement current \cite{blanter:2000}. This does not change our results. Imagine that a sample is screened electrostatically from the rest of the world. The drain is connected to a capacitor with a high-conductance wire.  Assume that we measure the displacement current in the capacitor. As long as the charge travels quickly along the wire and on the capacitor's plates, the low-frequency displacement current in the capacitor is the same as the charge current in the drain.

In conclusion, different edge-mode velocities in a spin-polarized FQHE state may result in excess 
auto-correlation noise. The effect is absent for cross-correlation noise, giving a more reliable tool for probing fractional charges.

\begin{acknowledgments}
We thank D. C. Glattli for sharing unpublished data with us and acknowledge useful discussions with D. C. Glattli and B. I. Halperin. This research was supported in part by the National Science Foundation under Grants No. DMR-2204635  and PHY-1748958.
\end{acknowledgments}

\end{document}